 \documentclass[final,3p,times,authoryear]{elsarticle}

\usepackage{amsmath,amssymb,amsfonts,hyperref,graphics,array,enumitem,booktabs,lineno,textcomp}

\usepackage{newtxmath}
\makeatletter
\re@DeclareMathSymbol{\varepsilon}{\mathord}{lettersA}{34}
\re@DeclareMathSymbol{\delta}{\mathord}{lettersA}{14}
\makeatother

\newcolumntype{A}{>{\centering\arraybackslash} m{0.07\textwidth} }
\newcolumntype{B}{>{\centering\arraybackslash} m{0.11\textwidth} }
\newcolumntype{E}{>{\centering\arraybackslash} m{0.13\textwidth} }
\newcolumntype{F}{>{\centering\arraybackslash} m{0.05\textwidth} }
\newcolumntype{C}{>{\centering\arraybackslash} m{0.08\textwidth} }
\newcolumntype{D}{>{\centering\arraybackslash} m{0.13\textwidth} }
\newcolumntype{G}{>{\centering\arraybackslash} m{0.22\textwidth} }

\journal{Icarus}

\begin{document}

\begin{frontmatter}



\title{Isotopically distinct terrestrial planets via local accretion}


\author{Jingyi Mah}\corref{corrauthor}
\cortext[corrauthor]{Corresponding author}
\ead{mahjingyi@gmail.com}
\author{Ramon Brasser}
\address{Earth Life Science Institute, Tokyo Institute of Technology, Ookayama, Meguro-ku, Tokyo 152-8550, Japan}

\begin{abstract}
Combining isotopic constraints from meteorite data with dynamical models of planet formation proves to be advantageous in identifying the best model for terrestrial planet formation. Prior studies have shown that the probability of reproducing the distinct isotopic compositions of the Earth and Mars for both classical and Grand Tack models is very low. In the framework of the Grand Tack model, for Mars to be isotopically different from the Earth, it had to form under very specific conditions. Here, we subjected a fairly new and unexplored model—the depleted disc model—to the test. It presupposes that the region in the inner protoplanetary disc from Mars’ orbit and beyond is depleted in mass such that Mars is left with insufficient material to grow to a larger size. Our aim is to test the whether the distinct isotopic compositions of the Earth and Mars are a natural outcome of this model. We found that the terrestrial planets accrete material mostly locally and have feeding zones that are sufficiently distinct. The Earth and Mars, and by extension, Venus, can have distinct isotopic compositions if there is an isotopic gradient in the terrestrial planet region of the protoplanetary disc. Our results suggest that the material in the inner Solar System most likely did not undergo substantial mixing that homogenised the potential isotopic gradient, in contrast to the Grand Tack model where the feeding zones of the terrestrial planets are nearly identical due to the mixing of material by Jupiter’s migration.

\end{abstract}




\end{frontmatter}

\section{Introduction}
Martian meteorites such as the shergottite-nakhlite-chassignite (SNC) meteorites, as well as NWA 7034, Tissint, ALH 84001, Zagami and many others, reveal that the Earth and Mars are not entirely made up of the same material. The martian meteorites exhibit isotopic anomalies that are distinct from the Earth in $\Delta^{17}$O \citep{clayton&mayeda1983,clayton&mayeda1996,franchietal1999,rubinetal2000,mittlefehldtetal2008,ageeetal2013,wittmannetal2015}, $\varepsilon^{50}$Ti \citep{trinquieretal2009,zhangetal2011,zhangetal2012}, $\varepsilon^{51}$V \citep{nielsenetal2014}, $\varepsilon^{53}$Cr, $\varepsilon^{54}$Cr \citep{shukolyukov&lugmair2006,trinquieretal2007,trinquieretal2008,qinetal2010a,qinetal2010b,yamashitaetal2010,larsenetal2011,petitatetal2011,yamakawa&yin2014}, $\varepsilon^{62}$Ni \citep{tang&dauphas2014}, $\varepsilon^{92}$Mo \citep{burkhardtetal2011}, and $\varepsilon^{142}$Nd \citep{kruijeretal2017b}. The $\varepsilon$ notation refers to the deviation in parts per ten thousand of the isotope normalised to a standard whereas $\Delta^{17}$O is defined as $\delta^{17}\rm{O}_{\rm{VSMOW}}-0.52\delta^{18}\rm{O}_{\rm{VSMOW}}$ where $\delta$ is the deviation in parts per thousand of the isotope normalised to a standard. 

The oxygen isotope system is useful for differentiating isotopic signatures originating from mass-independent (intrinsic heterogeneities in the primordial solar nebula) or mass-dependent planetary processes \citep{clayton&mayeda1983}. The fact that martian meteorites are significantly enhanced in $^{17}$O relative to $^{16}$O and $^{18}$O compared to the Earth implies that the Earth and its fellow neighbour were most likely assembled from components originating from different reservoirs,and possibly, different regions in the protoplanetary disc \citep{wanke&dreibus1988,wanke&dreibus1994,lodders2000,warren2011,brasseretal2017,carlsonetal2018}.

To identify the potential building blocks of the terrestrial planets, the isotopic anomalies of different groups of meteorites are utilised as constraints in mixing models. For Mars, earlier results yield a combination of  85\% H + 11\% CV + 4\% CI based on oxygen isotopes (\citealp{lodders&fegley1997}; H: a component of the ordinary chondrites (OC) group; CV/CI: components of carbonaceous chondrites (CC) group) and 45\% EC + 55\% OC (EC: enstatite chondrites; \citealp{sanloupetal1999}). \cite{tang&dauphas2014} subsequently show that the composition suggested by \cite{sanloupetal1999} is a good match when taking into account constraints from other isotopes such as titanium, chromium, nickel and molybdenum. \cite{brasseretal2018} also arrived at a consistent result (68$^{+0}_{-39}$\% EC + 32$^{+35}_{-0}$\% OC) when combining dynamical simulations with the mixing model. When differentiated meteorites are considered, a component that is isotopically similar to the angrite meteorites could potentially be a major component (55\% angrite + 36\% H + 9\% CI; \citealp{fitoussietal2016}) if the maximum fraction of EC permitted is 15\%. For Earth, the mixing models give 70\% EC + 21\% H + 5\% CV + 4\% CI \citep{lodders2000}, 91\% EC + 7\% OC + 2\% CC \citep{dauphasetal2014b}, 71\% EC + 24\% OC + 5\% CC \citep{dauphas2017}, while dynamical modelling arrives at 87\% EC + 13\% OC in the framework of the Grand Tack model \citep{brasseretal2017,wooetal2018}. The larger fraction of enstatite chondrites for the Earth is in contrast with the results for Mars.

On the other hand, it has also been shown that the current repository of known meteorites does not constitute the whole picture of the Earth’s building blocks (\citealp{drake&righter2002}; see \citealp{mezgeretal2020} for a recent review). Plots of Ti versus Cr \citep{trinquieretal2009,warren2011} and Ru versus Mo \citep{dauphasetal2014a,fischer-goeddeetal2015} show that the Earth is an end member, suggesting that there is a missing piece in the puzzle: an additional reservoir in the region closer to the Sun than where the ECs are thought to have formed that is characterised by an enrichment in {\it s}-process isotopes. The Earth should also have accreted some material from this reservoir to offset the {\it s}-process deficit signature of the other meteorites \citep{fischer-goedde&kleine2017,renderetal2017} as its isotopic anomalies cannot be explained were it to only have accreted material that was isotopically similar to EC, OC, and CC. The absence of samples from this {\it s}-process enriched reservoir is intriguing. It has been suggested that this have all been incorporated into the terrestrial planets \citep{drake&righter2002,burkhardtetal2011,burkhardtetal2016} and indeed, \cite{fischer-goeddeetal2020} may have found evidence of this reservoir in ancient terrestrial rocks that show positive $\varepsilon^{100}$Ru anomalies.

Comparing the outcomes from models of terrestrial planet formation with the constraints from isotopic anomalies of the Earth and Mars could provide fresh insights into the conundrum. \cite{chambers2001} looked at where the terrestrial planets sourced their building material, i.e., their feeding zones, in the framework of the Classical model, and found that there is a gradient in the feeding zones of the terrestrial planets from Mercury to Mars. This is expected because in the Classical model the terrestrial planets mostly grow by accreting material locally. However, a fundamental shortcoming plagues the model. It systematically produces planets that are several times the mass of Mars around 1.5 au. This spurred successive developments and refinements to the Classical model to form the terrestrial planets with the correct masses and orbits. \cite{hansen2009} found that growing the terrestrial planets from material concentrated in a narrow annulus between 0.7 au and 1.0 au is able to reproduce the mass-semimajor axis distribution of the terrestrial planets. The means to truncate the material was lacking, however.

\cite{walshetal2011} proposed a dramatic inward-then-outward migration of Jupiter and Saturn (known as the Grand Tack) as a means to sculpt the inner protoplanetary disc. The Grand Tack model correctly reproduces the masses and orbits of the terrestrial planets but its predictions of the isotopic compositions of the terrestrial planets are barely consistent with the data \citep{wooetal2018}. It was found that the migration of the gas giants as advocated by the Grand Tack model causes the material in the terrestrial planet region to become thoroughly mixed (see \citealp{carlsonetal2018} for a review) and thus results in feeding zones that are indistinguishable for all the terrestrial planets \citep{wooetal2018}. It is, however, still possible, though with a low probability, for Earth and Mars to be isotopically distinct if Mars began as a planetary embryo stranded in the asteroid belt region without sufficient mass to grow larger and then subsequently got scattered inwards to its current orbit \citep{brasseretal2017}. Yet this scenario has its own problems, not least of which is that it may not be possible to grow Mars-sized planets in the asteroid belt \citep{walsh&levison2019}. 

In this work, we test the predictions of the depleted disc model \citep{izidoroetal2014,izidoroetal2015,raymond&izidoro2017} on the feeding zones of the terrestrial planets. This model was based on the model suggested earlier by \cite{hansen2009}. Truncating material in the terrestrial planet region of the protoplanetary disc has proven successful in reproducing the orbital configuration of the terrestrial planets. The depleted disc model presupposes a depletion in mass in the region from and beyond the orbit of Mars, leaving insufficient mass for Mars to accrete and grow to a larger size. However, the mechanism or means to bring about a mass depletion in the inner disc is still not well-understood. A recent study by \cite{walsh&levison2019} demonstrated that it is possible to arrive at the starting conditions of the depleted disc model naturally via collisional grinding among planetesimals in the inner disc. They attempted to model the dynamics between gas and solids in the protoplanetary disc more realistically by starting with a disc of planetesimals, including the effects of collisional fragmentation, and allowing all bodies in the system to interact gravitationally. Their simulations were unsuccessful, however, at reproducing the mass difference between Earth and Mars, and it could be due to the limited combination of initial conditions that was explored.

Originally proposed as an alternative to the Grand Tack model, the depleted disc model is successful in reproducing the small mass of Mars and the low mass of the asteroid belt without invoking the dramatic migration of Jupiter and Saturn, provided the surface density of solids in the terrestrial region follows a rather steep power law function of the distance from the Sun ($r^{-5.5}$) and the gas giants are on their current orbits \citep{izidoroetal2015}. That the preferred orbital configuration of the gas giants is one that is similar to the modern day configuration is perhaps not coincidental. \cite{clementetal2018} found that Mars' low mass is reproduced if the giant planets underwent an epoch of instability very early in the Solar System's history, suggesting that the orbits of Jupiter and Saturn may have stopped evolving drastically when the terrestrial planets were forming. Without the migration of the gas giants stirring up the material in the terrestrial planet region and contaminating it with material from distant regions in the protoplanetary disc, the terrestrial planets could have localised feeding zones and by extension, different isotopic compositions. Given the potential of the depleted disc model, a thorough investigation into its dynamics and cosmochemical predictions is therefore warranted.


\section{Methodology}

\begin{figure*}[ht]
    \centering
	\includegraphics[width=0.6\textwidth]{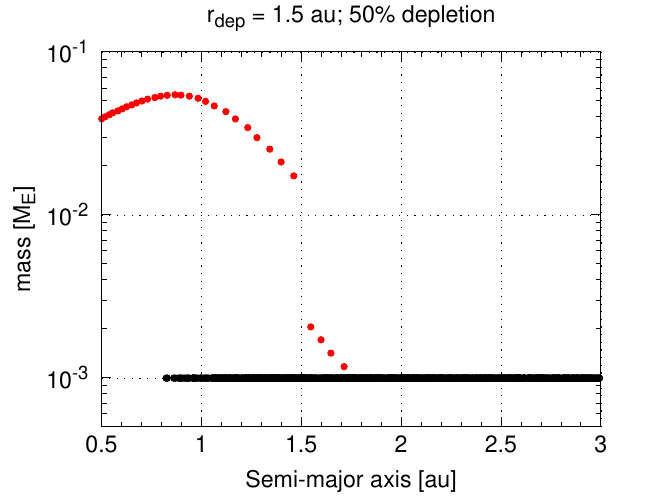}
    \caption{Initial masses (in units of Earth mass) and semi-major axis of planetary embryos (red dots) and planetesimals (black dots) for the initial condition $r_{\rm dep} =$ 1.5 au and 50\% mass depletion.}
    \label{fig:init_cond}
\end{figure*}

Our goal is to determine what the depleted disc model predicts for the feeding zones of the terrestrial planets and consequently, their final isotopic compositions. To this end, we first examine the dynamical outcomes and properties of planetary systems formed with this model, i.e., the resultant mass and semi-major axis distributions of planets, angular momentum deficit (AMD), mass concentration parameters, and spacing between the planets of the planetary systems. We test the performance of the model using N-body simulations with various initial conditions. Our variables are (i) the radii in the protoplanetary disc beyond where the mass is depleted $r_{\rm dep}$, and (ii) the scale of mass depletion. We tested for three values of $r_{\rm dep}$: 1.0 au, 1.25 au and 1.5 au, and three scales of mass depletion: 50\%, 75\%, and 95\%, for each $r_{\rm dep}$. This gave us a total of nine sets of initial conditions.

Our initial setup consists of a sequence of planetary embryos embedded in a disc of planetesimals (collectively referred to as solids). We set the inner edge of the solid disc to be at 0.5 au and the outer edge at 3.0 au. We followed the method of \cite{brasseretal2016} to generate the distribution of embryos and planetesimals, assuming that the embryos have undergone oligarchic growth \citep{kokubo&ida1998}. Our method employs the semi-analytical oligarchic approach of \cite{chambers2006}.  

As the first step, we computed the total mass in solids between 0.5 au and 3.0 au. We assumed a minimum-mass solar nebula (MMSN) solid surface density of $\Sigma_s =$ 7.1 g cm$^{-2}$ $(a/1 \ \rm{au})^{-3/2}$ \citep{hayashi1981} for the whole disc and scaled the solid surface density down by 50\%, 75\%, or 95\% in the region beyond the depletion radius $r_{\rm dep}$. The mass in the solid disc was then distributed into a sequence of feeding annuli for embryos spaced 10 mutual Hill radii apart following the results of \cite{kokubo&ida1998}. The initial spacing of the embryos is close to a geometric progression $a_n = a_{n-1}\Bigl[1 + b(2m_{\rm{iso}}/3M_{\odot})^{1/3}\Bigr]$, where $n$ is the index of the embryo, $b$ is the mutual spacing between the embryos, and $m_{\rm{iso}} = 2\pi a \Sigma_{s} b$ is the embryo isolation mass. The isolation mass is the maximum mass that the embryos can attain depending on their semi-major axes $a$, their mutual spacing $b$, and the surface density of the solid disc $\Sigma_s$. 

Next, we computed the masses of the embryos $m_p$ in their respective feeding annuli as a function of time according to \citep{chambers2006}
\begin{equation}
m_p(t)=m_{\rm{iso}}\tanh^3 \left( \frac{t}{\tau} \right).
\end{equation}
In the equation above, $\tau$ is the embryo growth timescale introduced in \cite{chambers2006}. It depends on the radii of the planetesimals $r_p$, in addition to $a$, $b$, and $\Sigma_s$. We assumed $r_p =$10 km when computing the embryos' growth timescales. The age of the solid disc when Jupiter has fully formed is denoted by $t$. We used $t = 1$ Myr in this work, based on the results of \cite{kruijeretal2017a}. We chose to conduct this study with only one value of $t$ to keep the number of simulations reasonable. At 1 Myr, the embryos would have attained only a fraction of their isolation masses. Finally, the remaining mass in the feeding annuli was subsequently allocated to planetesimals with masses equal to 0.001 Earth mass ($M_{\rm E}$) each. The leftover mass in the disc at this stage is distributed to planetesimals so that the solid disc extends to the outer edge at 3.0 au. 

The embryos and planetesimals were assigned orbital eccentricities $e$ and orbital inclinations $I$ from a Rayleigh distribution with scale parameters of the eccentricity distribution $\sigma_e = (m_p/3M_{\odot})^{1/3}$, and the inclination distribution $\sigma_I = 0.5\times\sigma_e$, respectively. The remaining orbital angles were chosen at random between $0^{\circ}$ and $360^{\circ}$ from a uniform distribution.

The evolution of systems of planetary embryos and planetesimals, including Jupiter and Saturn on their current orbits, were simulated using the SyMBA N-body integrator \citep{duncanetal1998} with a time step of 3.65 days. In our simulations, the gas giants and the embryos were able to gravitationally interact with each other and the planetesimals. The planetesimals however, were unable to interact with themselves. Bodies were removed from the simulations during collisions, in which case the bodies were assumed to have merged into a larger body perfectly, or when they ventured too close to ($a <$ 0.3 au) or too far from ($a >$ 100 au) the Sun. 

The simulations were executed in two phases. In the first 5 Myr, we included the effect of a gas disc that dissipates away in time following a power law function of the accretion rate onto the star. Specifically $\log \frac{\dot{M}_*}{M_\odot\,{\rm yr}^{-1}}= {-8} - \frac{7}{5}\log \Bigl( \frac{t}{1 {\rm Myr}}+0.1 \Bigr)$ \citep{hartmannetal1998,bitschetal2015}. 

The gas disc model we employed is based on the prescriptions of \cite{idaetal2016}. The disk midplane temperature is given by $T = \max (T_{\rm vis}, T_{\rm irr})$, where $T_{\rm vis}$ and $T_{\rm irr}$ are temperatures determined by viscous heating and stellar irradiation, respectively. Close to the star viscous heating dominates while far away stellar irradiation is the main heating source. We have 
\begin{align}
T_{\rm vis} &= T_{0v} \alpha_3^{-1/5} \dot{M}_{*8}^{2/5}
\left(\frac{r}{1\,{\rm AU}}\right)^{-9/10}\; {\rm K}, \nonumber \\
T_{\rm irr} &= 150 \left(\frac{r}{1\,{\rm AU}}\right)^{-3/7}\; {\rm K},
\label{eq:T_visirr}
\end{align}
where $r$ is the distance to the Sun and the power exponents are derived by analytical arguments. We also defined
\begin{align}
\alpha_3 &\equiv \frac{\alpha}{10^{-3}},  \; \; \\
\dot{M}_{*8} &\equiv \frac{\dot{M}_*}{10^{-8}\,M_\odot/{\rm yr}}, \; \;
\end{align}
where $\alpha$ is the viscosity \citep{shakura&sunyaev1973}.
With these temperature profiles, we compute the reduced scale height $h=H/r$ as
\begin{align}
h_{\rm vis} &= 0.034 \left(\frac{T_{0v}}{200\,{\rm K}} \right)^{1/2} \alpha_3^{-1/10}\dot{M}_{*8}^{1/5} 
\left(\frac{r}{1\,{\rm AU}}\right)^{1/20}, \nonumber \\
h_{\rm irr} &= 0.029 \left(\frac{r}{1\,{\rm AU}}\right)^{2/7}.
\label{eq:h_visirr}
\end{align}
The actual reduced scale height is given by $h = \max(h_{\rm vis},h_{\rm irr})$. The surface density of the gas is given by
\begin{align}
\Sigma_{\rm vis} &= 1320 \left(\frac{T_{0v}}{200\,{\rm K}} \right)^{-1} \alpha_3^{-4/5} \dot{M}_{*8}^{3/5}
\left(\frac{r}{1\,{\rm AU}}\right)^{-3/5}{\rm g\,cm}^{-2},\nonumber \\
\Sigma_{\rm irr} &= 1785 \alpha_3^{-1} \dot{M}_{*8}
\left(\frac{r}{1\,{\rm AU}}\right)^{-15/14}{\rm g\,cm}^{-2}.
\label{eq:Sigma_visirr}
\end{align}
The boundary between the viscous and irradiation regimes given by $T_{\rm vis} = T_{\rm irr}$ which occurs at
\begin{equation}
r_{\rm vis/irr} = \left(\frac{T_{0v}}{150\,{\rm K}} \right)^{70/33} \alpha_3^{-14/33}\dot{M}_{*8}^{28/33} \sim 1.84 \, 
\alpha_3^{-14/33}\dot{M}_{*8}^{28/33} {\rm au}.
\label{eq:r_vis_irr}
\end{equation}

The gas disk exerts torques and tidal forces on the embedded planets which result in a combined effect of radial migration and the damping of the eccentricity and inclination. The gas disc served to damp the eccentricities and inclinations of the embryos and planetesimals. We excluded the effect of gas-disc-induced migration (type-I migration) in all our simulations but retained the effects of eccentricity and inclination damping. As our focus in this work is on the isotopic composition of the terrestrial planets, we have chosen to simplify the approach while we reserve studying the effects of type-I migration for future work. Briefly, our prescription is as follows. The normalised torque is \citep{paardekooperetal2011}
\begin{equation}
 \frac{\gamma\Gamma}{\Gamma_0} = \frac{\Gamma_{\rm C}}{\Gamma_0}F_C + \frac{\Gamma_{\rm L}}{\Gamma_0}F_L
 \label{eq:torque}
\end{equation}
where $\Gamma_{\rm C}$ and  $\Gamma_{\rm L}$ are the corotation and 
Lindblad torques respectively and $\Gamma_0= (m_p/m_\odot)^2(H/r)^{-2}\Sigma\Omega_{\rm K}^2$ is a normalisation 
constant. The factors $F_L$ and $F_C$ are \citep{coleman&nelson2014}
\begin{eqnarray}
\ln F_C &=& -\frac{e}{e_{\rm f}}, \nonumber \\
\frac{1}{F_L} &=& P_e+{\rm sign} (P_e)(0.07\hat{i}+0.085\hat{i}^4-0.08\hat{e}^2\hat{i}^2),
\label{eq:fcfl}
\end{eqnarray}
where $e_{\rm f}= 0.01+{\textstyle \frac{1}{2}}h$, $\hat{e}=e/h$, $\hat{i} = \sin(i)/h$ and 
\begin{equation}
 P_e=\frac{1+(0.444\hat{e})^{1/2}+(0.352\hat{e})^6}{1-(0.495\hat{e})^4}.
\end{equation}
The eccentricity damping timescale $\tau_e=-e/\dot{e}$ is
\begin{equation}
\tau_{e} = 1.282t_{\rm wav}(1-0.14\hat{e}^2+0.06\hat{e}^3+0.18\hat{e}^2\hat{i}^2),
\label{eq:taue}
\end{equation}
where the wave timescale is \citep{tanaka&ward2004}.
\begin{equation}
t_{\rm wav} =\left(\frac{M_*}{m_p}\right)\left(\frac{M_*}{\Sigma r^2}\right)h^4 \Omega_K^{-1}.
\end{equation}
The inclination damping time scale $\tau_i = -i/(di/dt)$ is
\begin{equation}
\tau_{i} = 1.838t_{\rm wav}(1-0.30\hat{i}^2+0.24\hat{i}^3+0.14\hat{e}^2\hat{i}^2).
\end{equation}
After 5 Myr, we removed the gas disc artificially assuming that the gas has photoevaporated away completely by $t = 5$ Myr and then continued the simulations without the gas disc for another 150 Myr. 

As the evolution of each system is chaotic, a slight difference in the initial orbital configuration of each embryo and planetesimal will give different results. It is therefore necessary to perform many simulations to understand the range of possible outcomes. For each initial condition, we ran 16 simulations for a total of 144.

At the end of the simulations, we tabulated the number of terrestrial planets produced. We considered bodies with masses larger than 0.01 $M_{\rm E}$ to be planets. These planets were further filtered to identify good terrestrial planet analogues. We followed \cite{brasseretal2016} and imposed the following criteria for the masses and semi-major axes that the planets must comply with to qualify as good terrestrial planet analogues.
\begin{itemize}[noitemsep]
\item[--] Venus: 0.4 $M_{\rm E} < m_p <$ 1.2 $M_{\rm E}$, 0.55 au $ < a <$ 0.85 au
\item[--] Earth: 0.5 $M_{\rm E} < m_p <$ 1.5 $M_{\rm E}$, 0.85 au $ < a <$ 1.15 au
\item[--] Mars: 0.05 $M_{\rm E} < m_p <$ 0.15 $M_{\rm E}$, 1.3 au $ < a <$ 1.7 au
\end{itemize}
We did not attempt to place constraints for Mercury analogues as there remains many open questions about its formation history. 

For each good terrestrial planet analogue, we tracked their accretion histories and computed their feeding zones. This provides us with information on the region in the protoplanetary disc where the terrestrial planets sample most of their building blocks from. The feeding zone of a planet is quantified by the mean $a_{\rm{weight}}$ and width $\sigma_{a_{\rm{weight}}}$ of the initial semi-major axes of the solids accreted by the planet throughout its growth history weighted by their masses \citep{kaib&cowan2015}. These quantities are expressed mathematically as \citep{wooetal2018}
\begin{equation}
a_{\rm{weight}} = \frac{ \sum_{i}^{N} m_i a_i }{ \sum_{i}^{N} m_i },\ \rm{and}
\end{equation}

\begin{equation}
\sigma_{a_{\rm{weight}}} = \frac{ \sum_{i}^{N} m_i (a_i - a_{\rm{weight}})^2}{\frac{N-1}{N} \sum_{i}^{N} m_i },
\end{equation}
where $m_i$ and $a_i$ are the mass and semi-major axis of the \textit{i}th body accreted by the planet, and $N$ is the total number of bodies accreted. Together, $(a_{\rm{weight}}\pm\sigma_{a_{\rm{weight}}})$ define the feeding zone. The results are then compared with that obtained from the Grand Tack model \citep{brasseretal2016, wooetal2018}.


\section{Results and discussion}
\subsection{Architecture of terrestrial systems}
\label{sec:architecture}
\begin{figure*}[ht]
    \centering
	\includegraphics[width=0.9\textwidth]{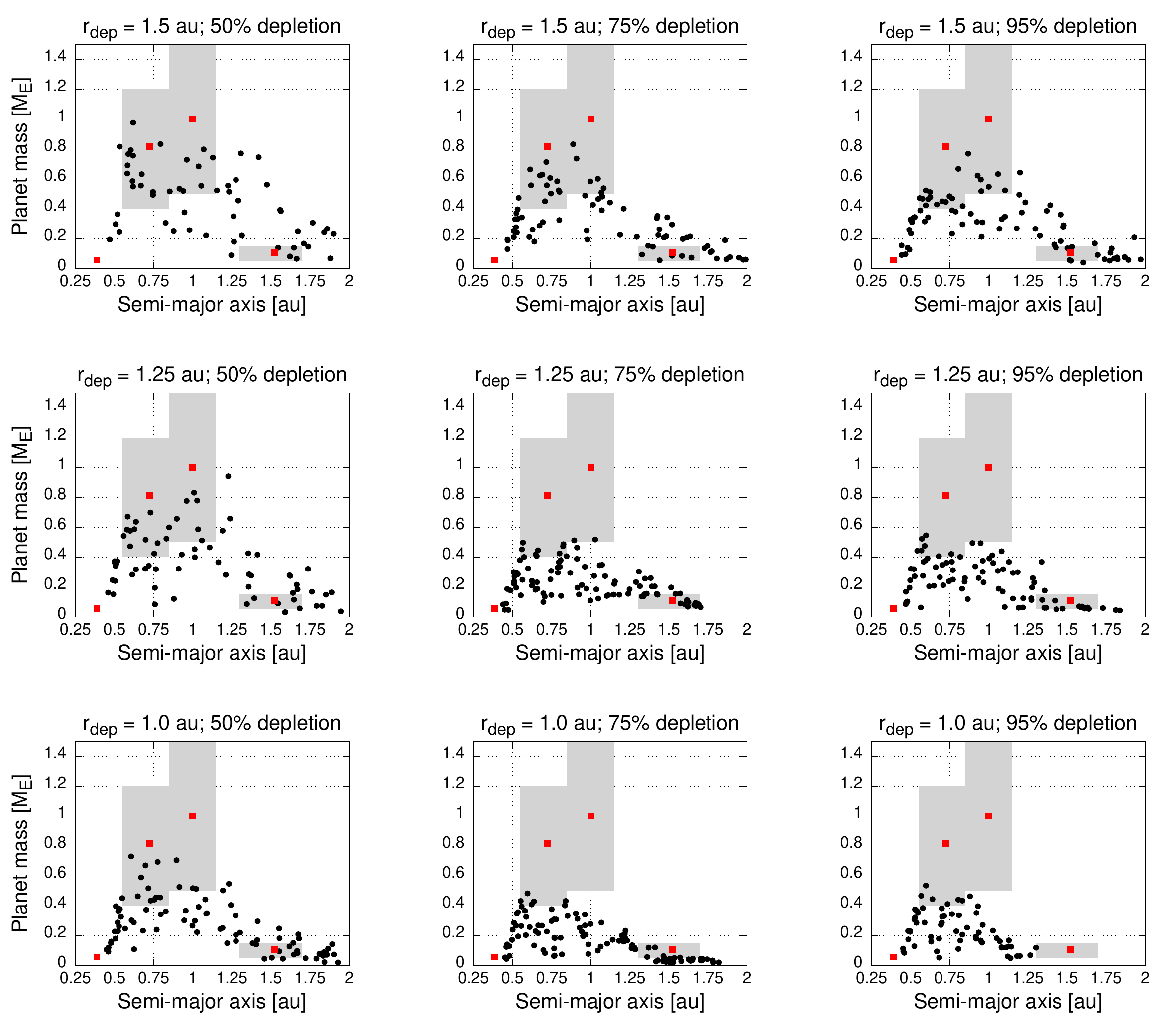}
    \caption{Mass and semi-major axis distribution of terrestrial planets formed at the end of our simulations for various scales of mass depletion at $r_{\rm dep} =$ 1.5 au (top panels), 1.25 au (middle row) and 1.0 au (bottom panels). Red squares are the current terrestrial planets, black circles are planets that formed in our simulations, and grey regions indicate the range within planets are considered good terrestrial analogues.}
    \label{fig:m_vs_a}
\end{figure*}

\begin{table*}[ht]
	\centering
	\caption{Total number of planets $N$, average number of planets in a system $\Bar{n}$, probability of forming Venus, Earth, and Mars analogues for each initial condition, and probability of yielding both Earth and Mars analogues in the same planetary system.}
    \label{tab:results}
    \begin{tabular}{AEFFBBBE}
    \toprule
    $r_{\rm dep}$ & Mass depletion scale & $N$ & $\Bar{n}$ & Venus analogues & Earth analogues & Mars analogues & Earth \& Mars analogues \\ 
    \midrule
    1.5 au  & 50\% & 57  & 3.6 & 23\% & 14\% & 7\% & 19\% \\
            & 75\% & 78  & 4.8 & 15\% &  8\% & 8\% & 13\% \\
            & 95\% & 80  & 5.0 & 18\% &  9\% & 8\% & 19\% \\
    \midrule     
    1.25 au & 50\% & 73  & 4.6 & 19\% & 8\% & 7\% & 19\% \\
            & 75\% & 94  & 5.9 & 9\% & 2\% & 16\% &  6\% \\
            & 95\% & 85  & 5.3 & 8\% & 0\% & 18\% &  0\% \\
    \midrule
    1.0 au  & 50\% & 86  & 5.4 & 12\% & 5\% & 9\% & 6\% \\
            & 75\% & 104 & 6.5 &  6\% & 0\% & 6\% & 0\% \\
            & 95\% & 63  & 3.9 & 10\% & 0\% & 2\% & 0\% \\
    \bottomrule
    \end{tabular}
\end{table*}

In Fig. \ref{fig:m_vs_a}, we show the mass and semi-major axis distribution of the terrestrial planets formed at the end of our simulations with varying mass depletion scales at depletion radius 1.5 au, 1.25 au and 1.0 au. The peak of the mass distribution is located between 0.75 au and 1 au for $r_{\rm dep} =$ 1.5 au (top panels of Fig. \ref{fig:m_vs_a}), although it is not clear for the case of 50\% mass depletion. The peak shifts Sunwards when the depletion radius $r_{\rm dep}$ is closer in. It is located at the current orbit of Venus for $r_{\rm dep} =$ 1.0 au (bottom panels of Fig. \ref{fig:m_vs_a}).

Most of the terrestrial planets produced in the simulations are generally less massive than the Earth and Venus, albeit with a few exceptions. Planet masses are also smaller when the depletion radius is closer to the Sun because the amount of mass available in the disc to form planets is smaller. This is an artifact of our initial conditions as we chose to keep the solid surface density fixed rather than increasing it in an attempt to reproduce the current masses of the terrestrial planets. Consequently, our simulations tend to produce more Venus analogues than Earth analogues (Table \ref{tab:results}) because of the low total solid mass in the disc. 

Our simulations produce planetary systems with an average of 4 to 6 planets. However, as there usually are planets within Venus' present orbit that could be considered as Mercury analogues, the higher number of planets poses no severe problems for the depleted disc model. From our simulation results, we find 52 planetary systems, out of a total of 144, with Mars analogues. Among these, only 16 have planets with $a >$ 1.7 au (upper limit of our criteria for Mars analogues) that could render them distinct from the Solar System.

We ran additional simulations with increased initial solid surface density of the solids in the protoplanetary disc and find that increasing the surface density by a factor of 1.5 to 2 times the MMSN value will produce planet analogues with masses closer to those of the current terrestrial planets. We discuss the outcomes in \ref{sec:appendix_mmsn}. 

Our results are consistent with that of \cite{walsh&levison2019} who also found for their model that the total mass in planets is lower than the current terrestrial system when the solid surface density value is that of the MMSN, although our initial conditions are different. Our starting configuration represents a later stage in time when embryos have started to form via the merging of planetesimals, and Jupiter and Saturn are fully formed and located at their current orbits. \cite{walsh&levison2019} opted to start with only planetesimals and the gas giants as Earth-mass embryos at semi-major axes 1.5 to 1.6 times closer to the Sun than their current values, representing an earlier stage of the protoplanetary disc. As such, a direct comparison between their results and ours is not possible, and only generalised statements will have to suffice.

\subsection{Dynamical properties of terrestrial systems}
\label{sec:dynamical_properties}
\begin{figure*}[ht]
    \centering
	\includegraphics[width=0.7\textwidth]{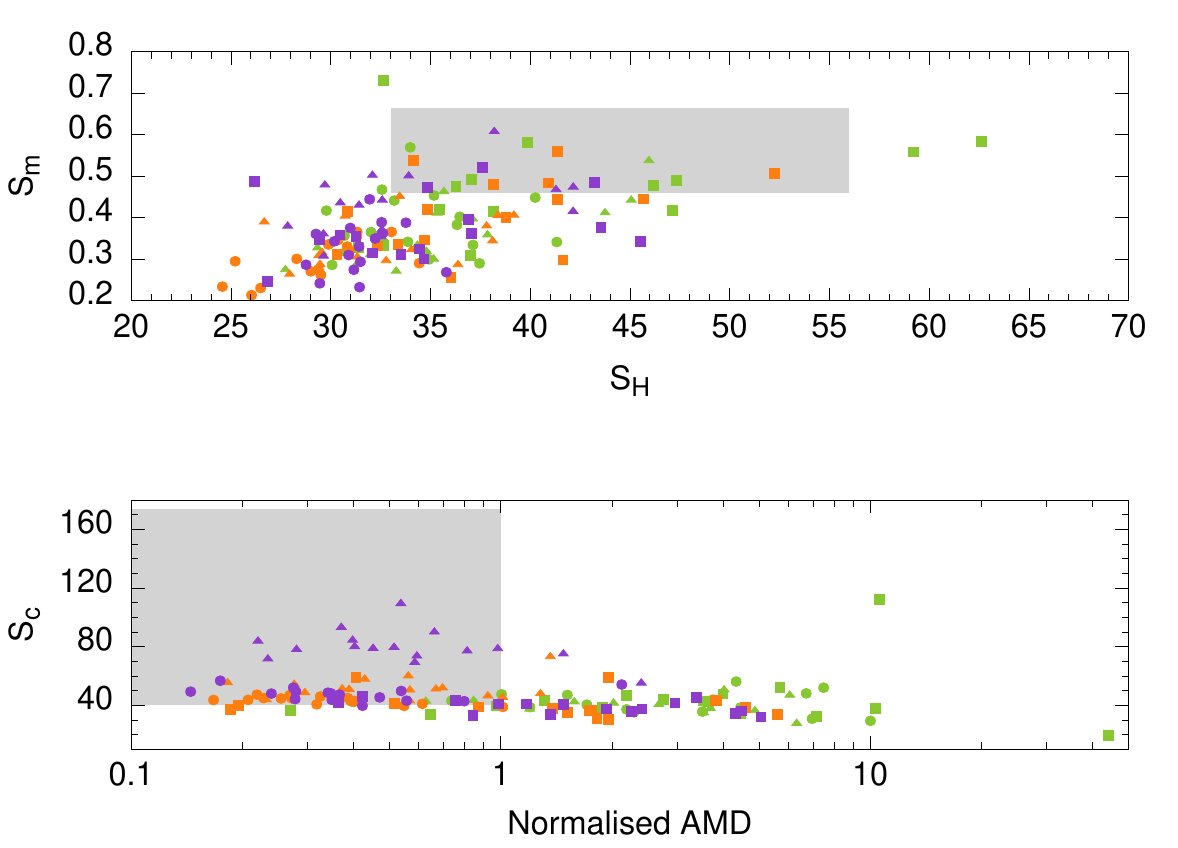}
    \caption{Top panel: fraction of total mass in the largest planet $S_m$ versus the mean spacing parameter $S_H$ in units of the mutual Hill sphere. Bottom panel: mass concentration statistic $S_c$ versus the angular momentum deficit (AMD) normalised to the current Solar System value. Green symbols represent simulations with $r_{\rm dep} = 1.5$ au, orange symbols represent $r_{\rm dep} = 1.25$ au, and purple symbols represent $r_{\rm dep} = 1.0$ au. Squares correspond to depletion scale of 50\%, circles to 75\%, and triangles to 95\%. The grey regions encompass the current values of the inner Solar System and the 2$\sigma$ range.}
    \label{fig:metrics}
\end{figure*}

We also examined the characteristics of final terrestrial systems produced from this model by employing several statistics introduced by \cite{chambers2001}. The first of these is the angular momentum deficit (AMD), defined as
\begin{equation}
    \rm{AMD} = \frac{\sum_k \mu_k \sqrt{a_k} \Bigl[ 1 - \sqrt{(1-e_k^2)} \cos{I_k} \Bigr]}{\sum_k \mu_k \sqrt{a_k}},
\end{equation}
where $\mu_k = m_k/M_{\odot}$. The second is the mass concentration parameter $S_c$ which measures the degree of mass concentration in one part of the planetary system, given by
\begin{equation}
    S_c = \rm{max} \left( \frac{\sum_k \mu_k}{\sum_k \mu_k \left( \log (a/a_k)^2 \right)} \right).
\end{equation}
The third is the fraction of total mass in the largest planet of the system $S_m$, and the last is the mean orbital spacing statistic $S_H$, given by
\begin{equation}
    S_H = \frac{2}{N-1}\sum^{N-1}_{k=1} \frac{a_{k+1}-a_k}{a_{k+1}+a_k} \left( \frac{\mu_{k+1}+\mu_k}{3} \right)^{-1/3}.
\end{equation}
We follow \cite{brasseretal2016} and use the mutual Hill sphere as the spacing unit for $S_H$.

The values of the aforementioned statistics for all of the planetary systems formed in our simulations are presented in Fig. \ref{fig:metrics}. We also plot the current values of the Solar System, and a 2$\sigma$ range obtained using a Monte Carlo method \citep{brasseretal2016}, in grey shaded areas.

In the top panel of Fig. \ref{fig:metrics}, we find that the majority of the planetary systems plot away and to the lower left part of the grey region, indicating that their $S_m$ and $S_H$ values are lower compared to the Solar System's current value. Their low $S_m$ and $S_H$ values mean that the mass difference between planets are small and the planets are more closely-packed than the current terrestrial planets. Only 20 out of a total of 144 planetary systems have similar $S_m$ and $S_H$ values to the current terrestrial system and most of them correspond to the initial condition of 50\% mass depletion (square symbols). Among these systems, 12 possess Earth analogues, 7 possess Mars analogues, and 6 possess both Earth and Mars analogues. There are 6 systems with neither Earth nor Mars analogues.

In terms of the concentration parameter $S_c$, the majority of the final planetary systems have low $S_c$ values but the initial condition of $r_{\rm dep} =$1.0 au and 95\% (purple triangles in the bottom panel of Fig. \ref{fig:metrics}) is able to produce planetary systems with higher $S_c$ values that are close to the Solar System's current value. However, these planetary systems failed to form any Earth analogues and barely succeed in forming Mars analogues. In terms of the AMD, the planetary systems have values that range widely from 10 times smaller to 10 times larger than the current value. We chose to define successful cases as planetary systems with AMD less than the current Solar System's value because the AMD is expected to increase in time due to chaotic diffusion \citep{laskar2008}. It is thus likely that the AMD at 4.5 Ga ago was lower than what it is today. These results of the dynamical properties of planetary systems formed in the framework of the depleted disc model are expected due to the fixed surface density that we employed.

\subsection{Feeding zones of terrestrial planets}
\label{sec:feeding_zones}
\begin{figure*}[ht]
    \centering
	\includegraphics[width=0.9\textwidth]{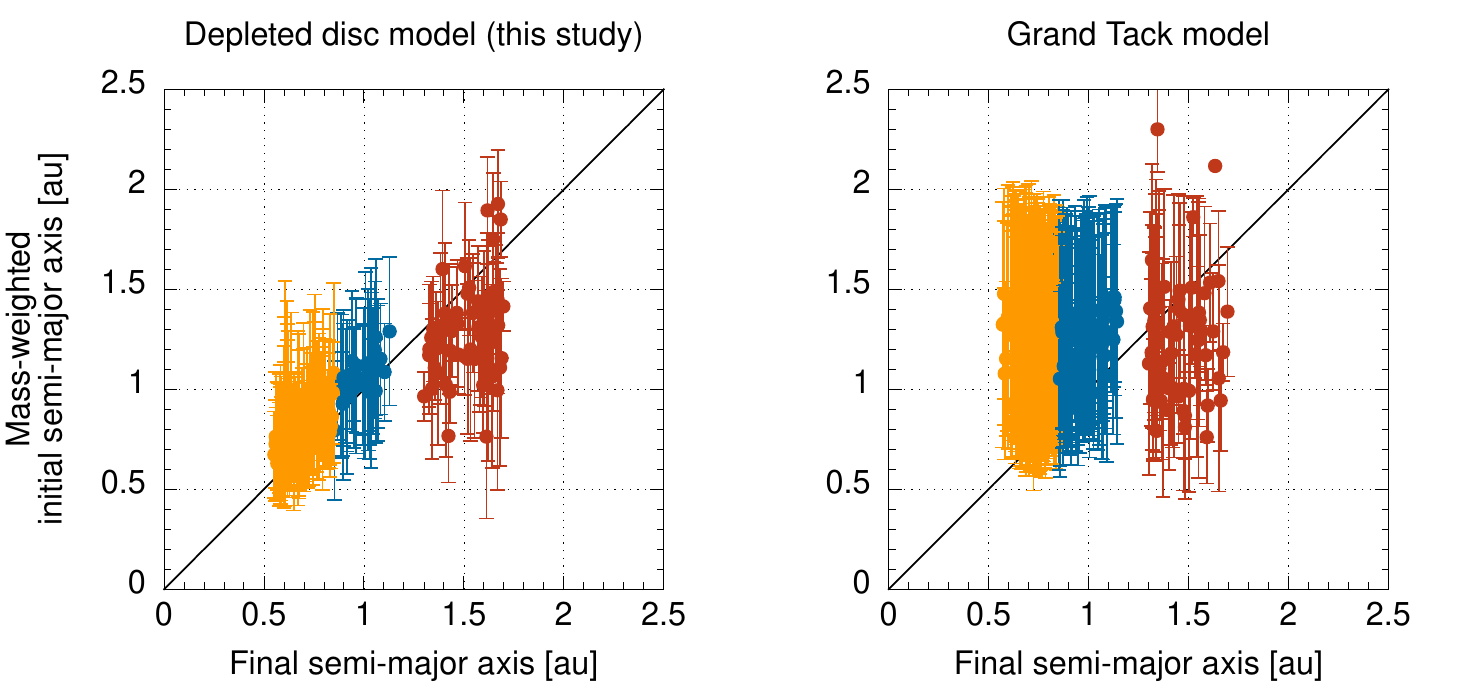}
    \caption{Feeding zones $(a_{\rm{weight}}\pm\sigma_{a_{\rm{weight}}})$ of the Venus (orange), Earth (blue) and Mars (red) analogues versus their final semi-major axes from the depleted disc model (left panel), and Grand Tack model (right panel) for comparison. Black line corresponds to the equation $y = x$. Feeding zone data for the Grand Tack model from \cite{wooetal2018}. The simulations used to compute the feeding zones are from \cite{brasseretal2016}. These include simulations with different tack locations for Jupiter (1.0 au and 1.5 au), simulations beginning with equal-mass planetary embryos, and simulations with embryos assumed to have underwent oligarchic growth. The results from all the aforementioned simulations are combined because \cite{brasseretal2016} found no obvious differences between the outcomes from different initial conditions.}
    \label{fig:acc_zone_compare}
\end{figure*}


\begin{figure*}[ht]
    \centering
	\includegraphics[width=0.6\textwidth]{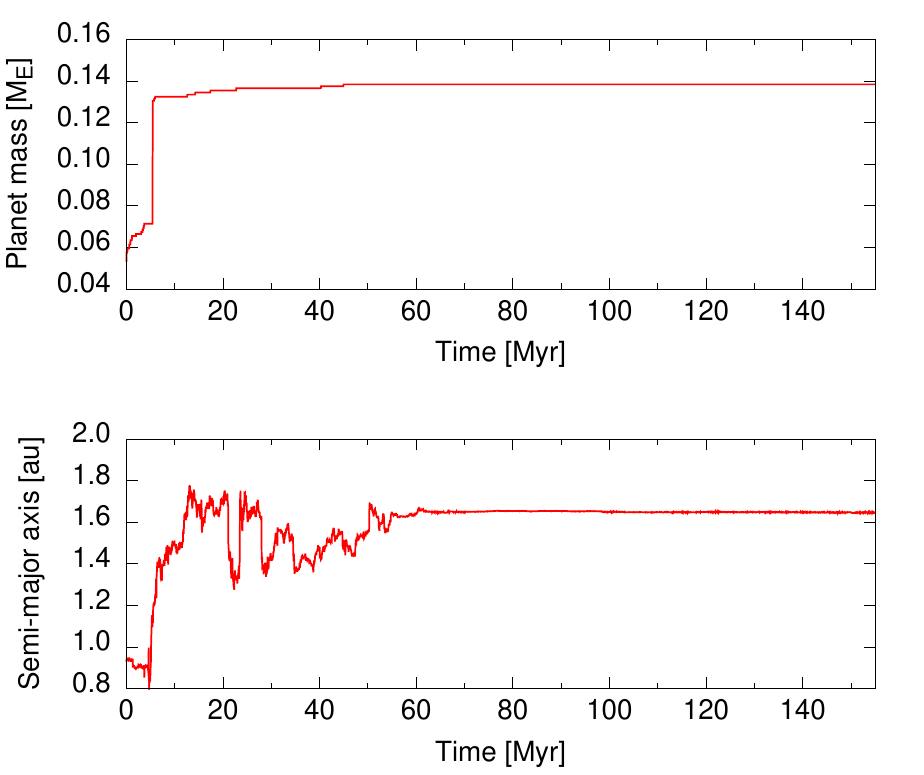}
    \caption{Time evolution of the mass and semi-major axis of a sample Mars analogue which feeding zone is centered at $a <$ 1.5 au.}
    \label{fig:mars_analog}
\end{figure*}

\begin{figure*}[ht]
    \centering
	\includegraphics[width=0.6\textwidth]{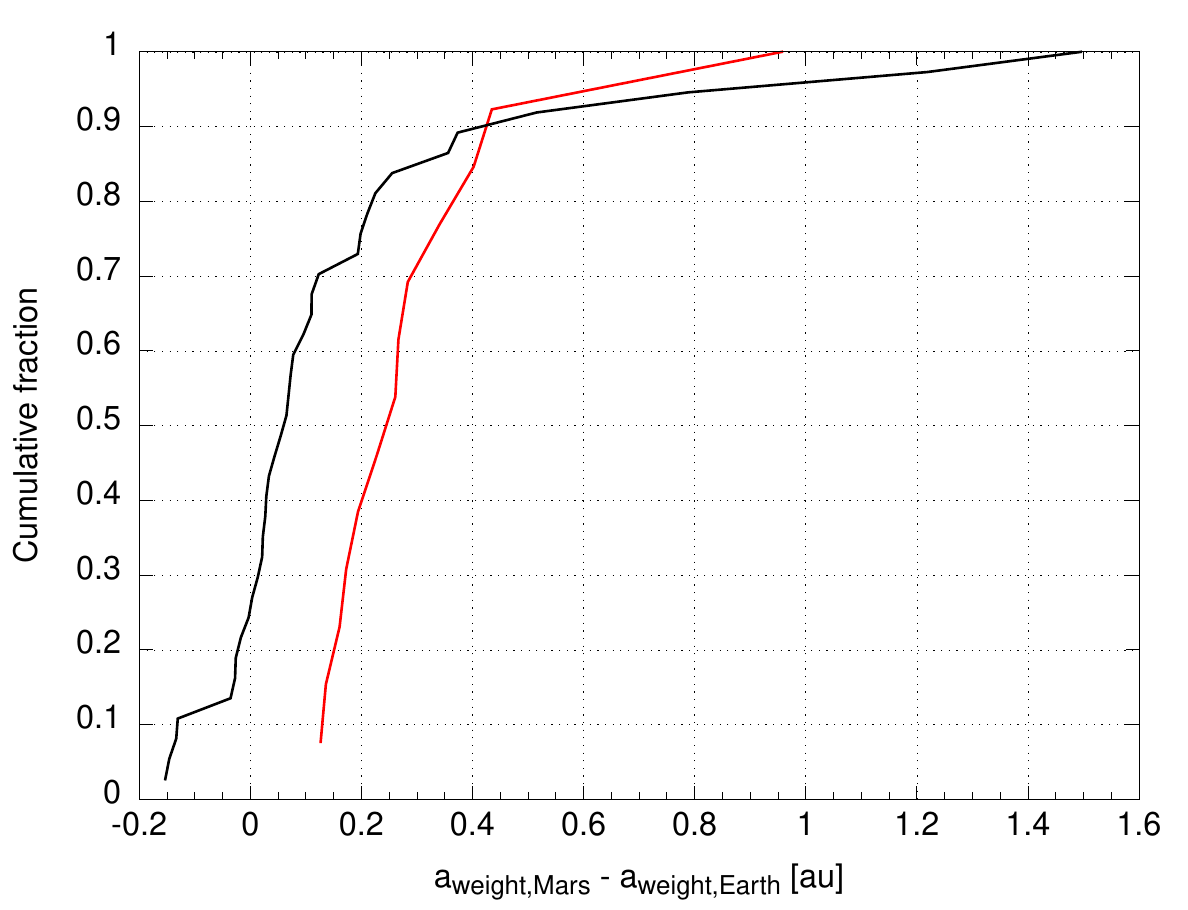}
    \caption{Cumulative distribution of the difference between the mass-weighted mean initial semi-major axis of solids (embryos and planetesimals) accreted by both Mars and Earth analogues in the depleted disc model (red) and the Grand Tack model (black).}
    \label{fig:aweight_compare}
\end{figure*}

\begin{figure*}[ht]
    \centering
	\includegraphics[width=0.7\textwidth]{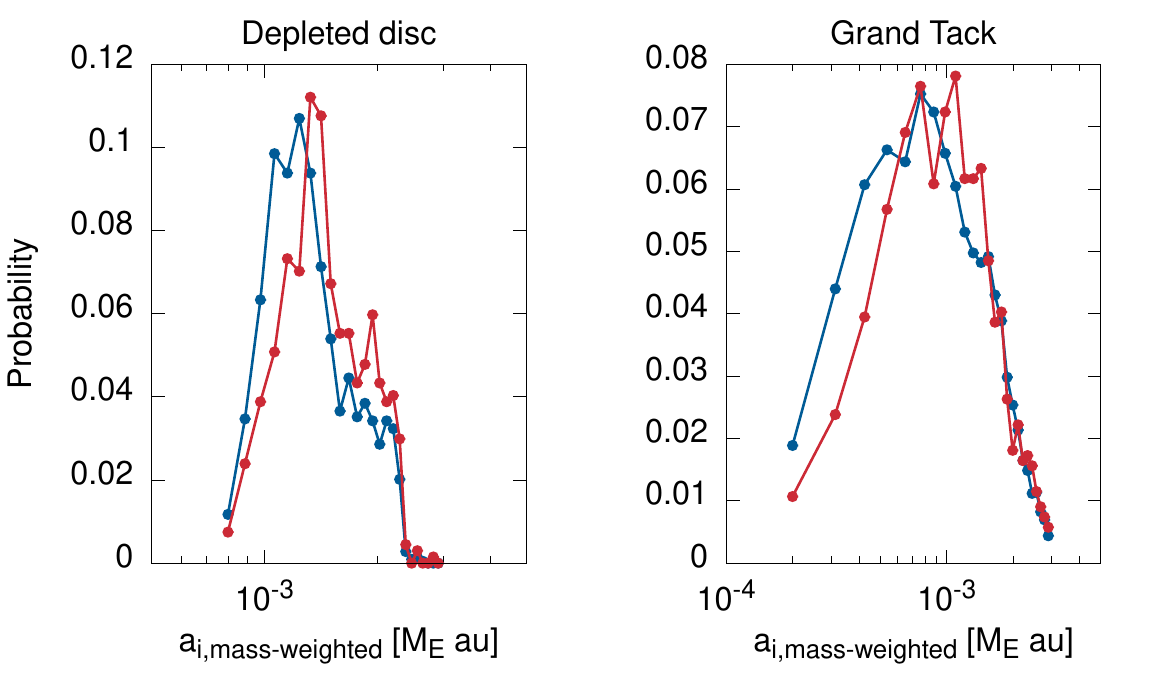}
    \caption{Probability distribution function of the mass-weighted initial semi-major axis of all the planetary embryos and planetesimals accreted by Earth (blue) and Mars (red) in the depleted disc model and the Grand Tack model. The distribution for Mars is systematically displaced to the right for the depleted disc model.}
    \label{fig:ma_compare}
\end{figure*}

\begin{figure*}[ht]
    \centering
	\includegraphics[width=\textwidth]{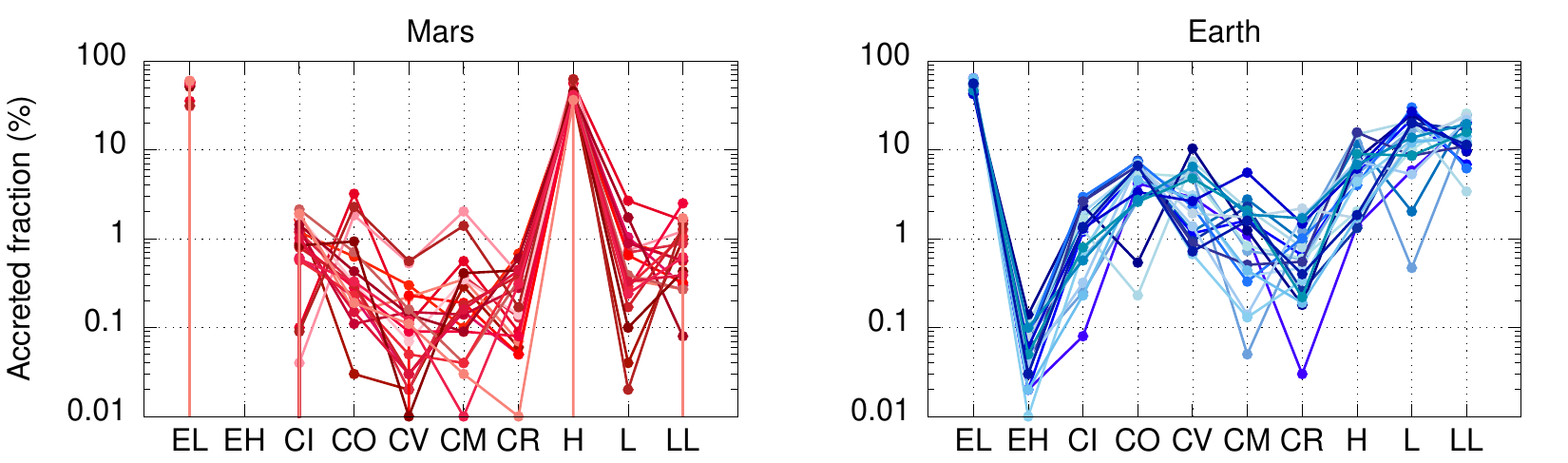}
    \caption{Contribution of each chondrite group to Mars and Earth obtained from a Monte Carlo mixing model based on isotope data.}
    \label{fig:mcmc_isocomp}
\end{figure*}

We present the feeding zones $(a_{\rm weight }\pm\sigma_{a_{\rm weight }})$ of the good terrestrial planet analogues, i.e., those that satisfy the criteria listed in the Methodology section, in the left panel of Fig. \ref{fig:acc_zone_compare}. There is a general increase in the heliocentric distance of the feeding zones with respect to the final semi-major axis of the terrestrial analogues, indicating that the planets tend to accrete material mostly locally. The trend of the local feeding zones likely arose from the eccentricity damping effect exerted by the gas on the embryos and planetesimals. Planetesimals are kept at low eccentricity (nearly circular) orbits and as a result they tend to get accreted onto nearby embryos. Embryos are typically 0.1 $M_{\rm E}$ or smaller so the escape velocity at their surfaces is roughly 5 km/s and thus the maximum eccentricity the embryos can obtain from embryo-embryo scattering is about 0.1. Therefore, the feeding zones for the final Earth and Venus analogues should be narrow and distinct.

The trend of the feeding zones does not include all Mars analogues, however. We observe a spread in the feeding zones of our Mars analogues and some of them deviate from the trend. We examined the evolution of these Mars analogues that deviate from the trend and found that they were formed closer in ($a <$ 1.5 au) and subsequently scattered outwards to their final orbits (Fig. \ref{fig:mars_analog}). Since most of their building blocks comprise material from within 1.5 au, it is reflected in their feeding zones that are centred at $a \approx$ 1.25 au.

The trend of the feeding zones of the terrestrial planets in the framework of the depleted disc model comports better with planetary isotopic compositions. In this model, Earth and Mars would accrete material from different regions in the protoplanetary disc. If these regions are isotopically different, then Earth and Mars can end up with different final mixtures of building material. In contrast, the feeding zones of the terrestrial planets in the Grand Tack model are wide and display no trend (right panel of Fig. \ref{fig:acc_zone_compare}; \cite{wooetal2018}). The migration of Jupiter and Saturn as proposed in the Grand Tack model results in the material in the region within Jupiter’s orbit being mixed \citep{brasseretal2018}. Consequently, the isotopic compositions of Earth and Mars should also be identical, in contradiction with isotopic data, unless a rather specific formation pathway for Mars is invoked \citep{brasseretal2017} . 

We expect the same outcome for the model proposed by \cite{walsh&levison2019} as their work demonstrates that the growth of planetary embryos from a disc of planetesimals is inside out and localised. The earliest embryos emerge in the inner region of the protoplanetary disc close to the Sun, quickly depleting the planetesimals in their vicinity while the planetesimals in outer region of the disc remain unperturbed because the embryo growth timescale in the outer region is longer. The embryo ``front'' then propagates outward with time but the embryos that grow further away from the Sun has lower mass because more mass is loss via collisional fragmentation with increasing distance from the Sun.

How different are the feeding zones of the Mars analogues compared to the Earth analogues in the depleted disc model and the Grand Tack model? We singled out the planetary systems which possess both Earth and Mars analogues (13 from the depleted disc model, 37 from the Grand Tack model) and computed the difference between $a_{\rm{weight, Mars}}$ and $a_{\rm{weight, Earth}}$. For systems with multiple Mars analogues, we chose the planet with semi-major axis closest to 1.5 au. The cumulative distribution is presented in Fig. \ref{fig:aweight_compare}. We find that the region in the disc where Mars analogues sourced most of their building blocks is generally more distant than the Earth in the depleted disc model, whereas it is closer to the Earth in the case of the Grand Tack model.

To further quantify the difference in the feeding zones of the Earth and Mars analogues produced in the depleted disc and Grand Tack models, we computed their overlapping coefficient (OVL), which is defined as the common area under two probability density functions. The OVL measures the similarity between two distributions. For planetary systems with both Earth and Mars analogues, we traced the accretion histories of the Earth and Mars analogues to obtain the mass-weighted initial semi-major axis of all the planetary embryos and planetesimals accreted onto these two planets. We then combined the mass-weighted initial semi-major axis data for all the Earth and Mars analogues to compute the OVL. This is done for better statistics as some of the Mars analogues only accreted a few planetesimals, which results in a rather grainy probability function. The OVL is computed using 
\begin{equation}
    {\rm OVL} = \sum_{m a} \min \Bigl[ f_{\rm Mars} ( m a ),f_{\rm Earth} ( m a ) \Bigr].
\end{equation}
We find ${\rm OVL}_{\rm DD} = 0.79$ for the depleted disc model, and ${\rm OVL}_{\rm GT} = 0.89$ for the Grand Tack model.

The OVL results suggest that the difference in feeding zones between Earth and Mars is $\sim20\%$ in the depleted disc model, and $\sim10\%$ in the Grand Tack model. What does this difference actually mean? Naively, one would expect a 10\% difference to imply that Mars' oxygen isotopes would result in a $\Delta^{17}$O of either +0.10 or -0.10\textperthousand; instead the observed anomaly of +0.29\textperthousand\ would suggest an OVL difference of $\sim30\%$, which is closer to the depleted disc value rather than the Grand Tack value. The question then becomes: What did Mars accrete that the Earth did not?

One interpretation of the different OVL values could be in the fraction of ordinary chondrites accreted by Mars because it is the only quantity that is potentially significantly different when one considers the terrestrial planets to be mixtures of chondrites \citep{lodders&fegley1997,sanloupetal1999,dauphas2017,brasseretal2018}. We have expanded on the mixing models of \cite{dauphas2017} and \cite{brasseretal2018} by adding several different meteorites groups of the carbonaceous chondrite variety (CM and CR; we split CO from CV) and splitting the enstatite and ordinary chondrites into their respective groups (EL and EH for enstatite chondrites, and L, LL and H for ordinary chondrites). Results of this updated mixing model are presented in Fig. \ref{fig:mcmc_isocomp} which shows the mixing models for the Earth and Mars respectively. We can apply the same technique to compute the OVL for the distributions in this mixing model. We ran 20 different iterations of the model for each planet and then computed the OVL averaging over all possible permutations between the two planets. We find ${\rm OVL}_{\rm MCMC}$ = 0.58$\pm$0.08, which is clearly attributed to the increased H chondrite fraction in Mars versus a higher fraction of carbonaceous material in the Earth. The typical compositions agree within uncertainty with those of \cite{brasseretal2018}. Taking the mixing model as our benchmark for the OVL, the depleted disc model does a better job than the Grand Tack.

\subsection{Relations to cosmochemistry, and implications for Venus}
\begin{figure*}[ht]
    \centering
	\includegraphics[width=\textwidth]{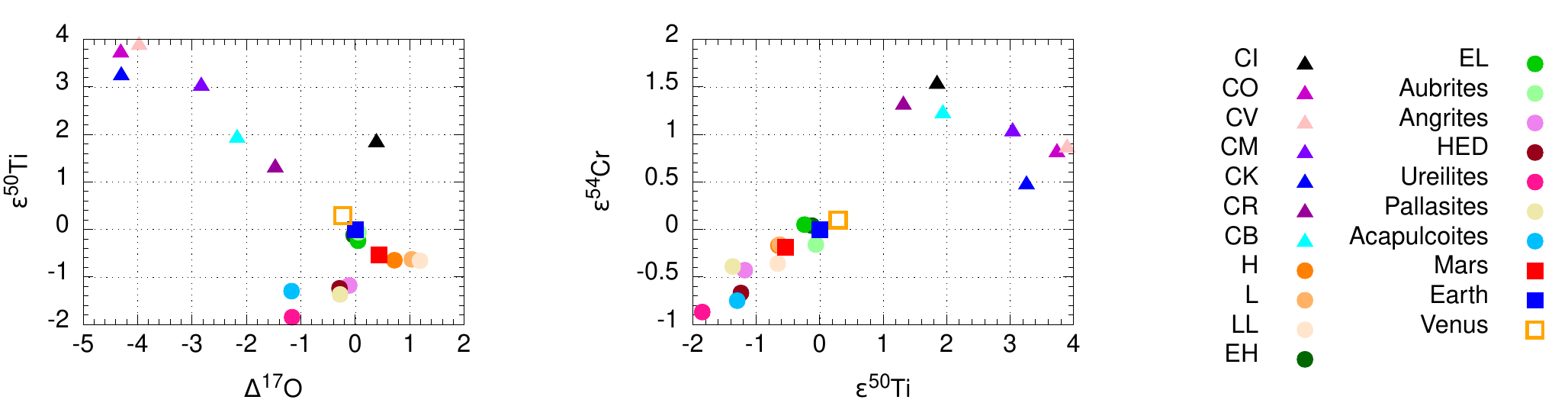}
    \caption{Isotopic anomalies in oxygen, titanium, and chromium for meteorites, Earth, and Mars. Meteorites in the carbonaceous/jovian group are labelled as triangles whereas meteorites in the non-carbonaceous/terrestrial group are labelled as circles. The terrestrial planets and our prediction for Venus are labelled as squares. Data from the compilation of \cite{brasser&mojzsis2020} and references therein. Martian meteorites used to compute the average value for Mars are ALH 84001, DaG 476, Chassigny, Lafayette, Nakhla, NWA 856, NWA 2737, NWA 7034, Shergotty and Zagami.}
    \label{fig:isotopes}
\end{figure*}

Fig. \ref{fig:isotopes} is a compilation of the isotopic variations in $^{17}$O \citep{clayton&mayeda1983,clayton&mayeda1996,franchietal1999,rubinetal2000,mittlefehldtetal2008,ageeetal2013,wittmannetal2015}, $^{50}$Ti \citep{trinquieretal2009,zhangetal2011,zhangetal2012}, and $^{54}$Cr \citep{shukolyukov&lugmair2006,trinquieretal2007,trinquieretal2008,qinetal2010a,qinetal2010b,yamashitaetal2010,larsenetal2011,petitatetal2011,yamakawa&yin2014} that have been reported in the literature for various meteorites. The meteorites fall into two distinct groups: (i) the non-carbonaceous/terrestrial group which consists of meteorites and differentiated bodies that are thought to have formed in the warmer regions of the protoplanetary disc close to the Sun, and (ii) the carbonaceous/jovian group which constituents are considered to have formed in the distant regions of the protoplanetary disc \citep{warren2011}. In addition to the dichotomy, it is clear that some nucelosynthetic isotopic tracers have strong correlations with each other, while others are more scattered \citep{warren2011}.

Several prior studies discussed a trend in the isotopic variations of titanium \citep{trinquieretal2009}, chromium \citep{trinquieretal2008,yamakawaetal2010}, nickel \citep{regelousetal2008}, molybdenum \citep{renderetal2017}, ruthenium \citep{fischer-goeddeetal2015,fischer-goedde&kleine2017}, and neodymium \citep{bouvier&boyet2016,burkhardtetal2016} among various meteorite groups. Specifically, \cite{yamakawaetal2010} suggested that the differences in $\varepsilon^{54}$Cr of the Earth, Mars and Vesta implied a gradient in the isotopic composition of the protoplanetary disk. The distinct isotopic compositions of Earth and Mars support the existence of such a gradient on the basis that the bulk isotopic compositions of planetary bodies should reflect the composition of the disc from which they sourced their building blocks. If the idea proposed by \cite{yamakawaetal2010} is correct then Vesta should have formed near its current position, in a region where \cite{walsh&levison2019} claim it is difficult to grow large bodies.

The distinct isotopic compositions of Earth and Mars could suggest that their feeding zones did not greatly overlap. Our simulation results show that Venus, Earth and Mars accreted from different, localised regions of the protoplanetary disk. Thus, a direct implication of the depleted disc model is that Venus should be isotopically distinct from the Earth and Mars. A prediction of its position in Fig. \ref{fig:isotopes} can be obtained by making use of the correlation between the feeding zones of the planets shown in Fig. \ref{fig:acc_zone_compare} and the distinct isotopic compositions of the Earth and Mars in $\Delta^{17}$O, $\varepsilon^{50}$Ti and $\varepsilon^{54}$Cr. This assumes that the disc's isotopic composition varies linearly with semi-major axis \citep{pahlevan&stevenson2007}. The predicted isotopic anomalies for Venus $\varepsilon_{\rm V}$ can be computed using the relation
\begin{equation}
\label{eq:isotope}
    \varepsilon_{\rm V} = \frac{a_{\rm V}-a_{\rm E}}{a_{\rm M}-a_{\rm E}} \times \varepsilon_{\rm M},
\end{equation}
where the subscripts V, E and M refer to Venus, Earth and Mars, respectively. We computed the nominal values of the isotopic anomalies of Venus using the current semi-major axes of the terrestrial planets and plotted them as open squares in of Fig. \ref{fig:isotopes}. 

In computing the nominal isotopic composition for Venus, we used the Earth-Mars correlation instead of Earth-Mars-Vesta. Although the Earth-Mars-Vesta trend is present for $\varepsilon^{50}$Ti and $\varepsilon^{54}$Cr, it is not the case for $\Delta^{17}$O where Vesta has a negative value, in contrast to the positive values of the Earth and Mars. Furthermore, as we only focused on the feeding zones of the terrestrial planets in this work, we opted to be consistent and thus exclude Vesta in our computations.

The uncertainties in the isotopic anomalies were computed using a Monte Carlo method. We employed the Box-Mueller transform to generate values for the semi-major axes of Venus, Earth, and Mars, as well as the isotopic anomalies in $\Delta^{17}$O, $\varepsilon^{50}$Ti and $\varepsilon^{54}$Cr for Mars according to a normal distribution with mean equal to the mean semi-major axes of the feeding zones of the planets, and the standard deviation being a quarter of the feeding zones' width. We then computed the isotopic anomalies of Venus using Eq. \ref{eq:isotope}. After 10$^5$ iterations, we obtain the isotopic anomalies and their corresponding uncertainties by computing the mean, 5th percentile and 95th percentile values. We report the values in Table \ref{tab:isotopes}. There is a huge spread in the predicted isotopic anomalies of Venus, and this is due to the fact that (i) the value of $(a_{\rm V}-a_{\rm E})/(a_{\rm M}-a_{\rm E})$ can be very small, which results in large variations, and (ii) some Mars analogues do not follow the same trend as the Earth and Venus, which skews the values. We also find that about half of the Venus analogues generated by the Monte Carlo method are predicted to have isotopic compositions that are more distinct from the Earth compared to Mars.


\begin{table*}[ht]
	\centering
	\caption{Predicted values for the isotopic anomalies of Venus by linear extrapolation of the Earth-Mars trend, and by using a Monte Carlo method, and the fraction of Venus analogues (generated by the Monte Carlo method) that is more isotopically distinct from the Earth than Mars is.}
    \label{tab:isotopes}
    \begin{tabular}{CDDG}
    \toprule
                & Nominal & Monte Carlo & $\lvert \varepsilon_V-\varepsilon_E \rvert > \lvert \varepsilon_E-\varepsilon_M \rvert$\\ 
    \midrule
    $\Delta^{17}$O       & -0.23 & -0.12$^{+2.69}_{-2.74}$ & 50\%\\[1.6ex]
    $\varepsilon^{50}$Ti & +0.29 & +0.26$^{+3.37}_{-3.43}$ & 48\%\\[1.6ex]
    $\varepsilon^{54}$Cr & +0.10 & +0.07$^{+1.20}_{-1.19}$ & 49\%\\
    \bottomrule
    \end{tabular}
\end{table*}

Interestingly enough, using the extrapolation from the Earth and Mars trend implies that Venus plots in between the Earth and the carbonaceous/jovian material, which could naively be interpreted as Venus having accreted a higher fraction of carbonaceous/jovian material than the Earth, although this interpretation makes little sense dynamically. The Earth-Mars-Vesta trend in $\varepsilon^{50}$Ti and $\varepsilon^{54}$Cr is continued to Venus, but since Vesta is negative in $\Delta^{17}$O there is no such trend for the oxygen isotopes. Our predicted isotopic compositions for Venus also suggest that the angrite meteorites could not have originated from Venus.

Isotope measurements for meteorites or rock samples from Venus will therefore be crucial to know its oxygen isotopic composition \citep{greenwood&anand2020}, and provide evidence for or against the presence of a reservoir enriched in {\it s}-process elements that the Earth is said to have accreted some of its building blocks from \citep{renderetal2017}. Venus should have accreted a larger fraction of its building blocks from this {\it s}-process nuclides enriched reservoir given the presence of a potential isotopic gradient in the protoplanetary disc and that it is located closer to the Sun as compared to the Earth.


\section{Conclusions}
We examined the masses, orbital configurations, and feeding zones of the terrestrial planets formed in the framework of the depleted disc model by running a large number of N-body simulations with different initial conditions and including the effects of a dissipating gas disc during the first few Myr of the simulations. We found that the model outputs planets that are less massive than the current terrestrial planets if we assume a MMSN surface density for the solids in the protoplanetary disc, but the model is successful nevertheless in producing planets with low mass in the region near Mars’ orbit. Increasing the initial surface density of the solids in the protoplanetary disc by 1.5 to 2 times the MMSN value will resolve the deficit in the terrestrial planets' mass. As for the feeding zones of the terrestrial planets, our results show that the terrestrial planets accrete mostly locally. The trend of distinct feeding zones for each terrestrial planet can explain why the Earth and Mars are isotopically different.  

Despite the mass depletion in the protoplanetary disc being an ad-hoc assumption, the depleted disc model provides a promising initial condition to form Mars with the correct mass, a problem that has plagued the classical model and inspired the development of subsequent models. Perhaps more importantly, this model predicts isotopic compositions for the Earth and Mars that are distinct and consistent with isotope data, suggesting that the material in the region of the protoplanetary disc where the terrestrial planets were growing most likely did not experience thorough mixing that homogenised a putative isotopic gradient.


\appendix
\section{Effect of increasing the initial solid surface density}
\label{sec:appendix_mmsn}
Planetary systems with total mass lower than the current terrestrial system are expected outcomes given our choice to adopt a fixed initial surface density of solids in the protoplanetary disc. Increasing the initial surface density should reproduce the masses of the terrestrial planets. Here we report our results of the additional simulations we ran for $r_{\rm dep} =$ 1.5 au and 1.0 au with initial surface density 1.5 to 2 times the MMSN value. 

\begin{figure*}[ht]
    \centering
	\includegraphics[width=0.9\textwidth]{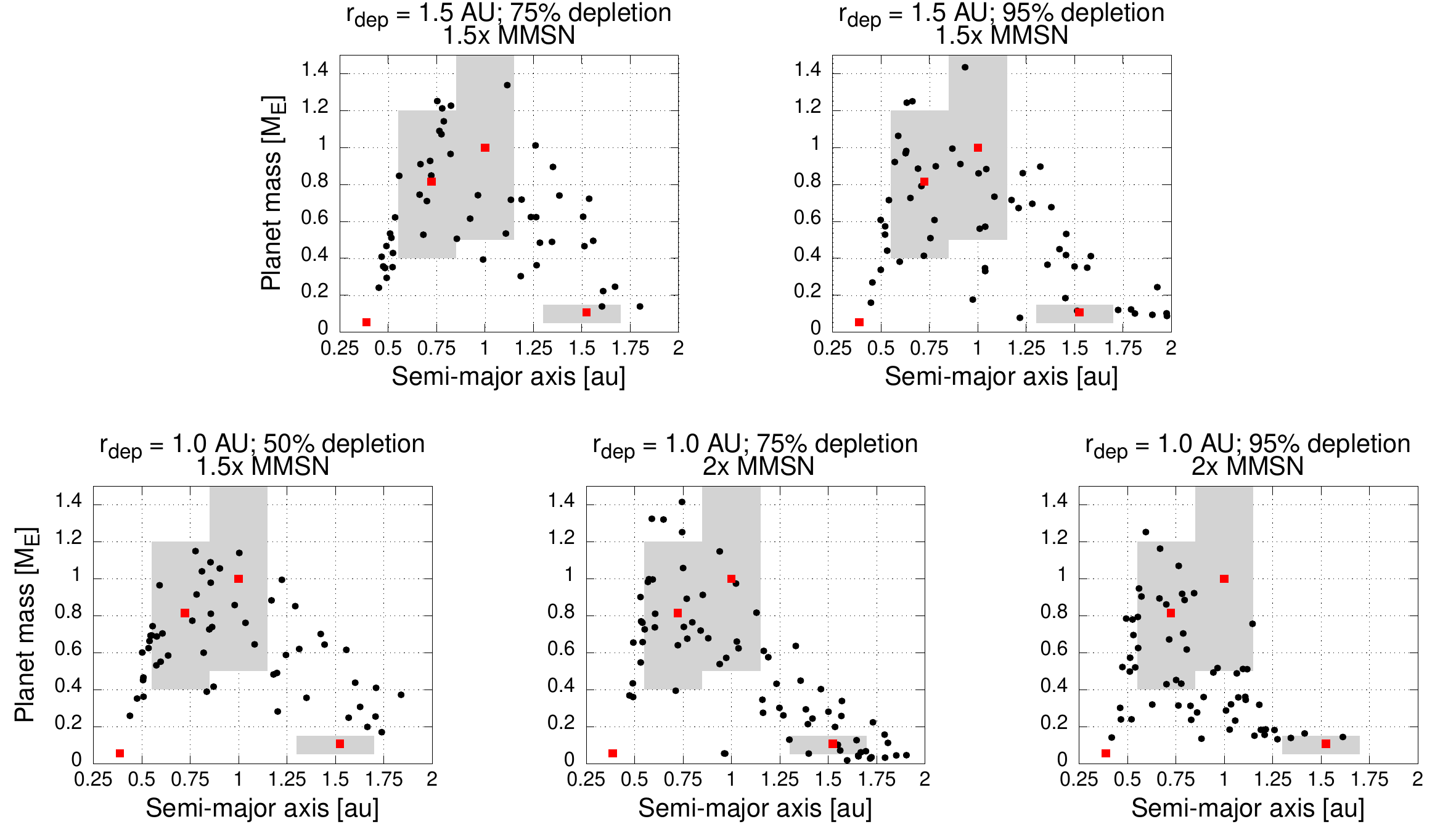}
    \caption{Results of increasing the initial surface density of solids in the protoplanetary disc: Mass and semi-major axis distribution of planets.}
    \label{fig:m_vs_a_extra}
\end{figure*}

Fig. \ref{fig:m_vs_a_extra} shows the distribution of mass versus semi-major axis of the final planets (bodies with mass $> 0.01 M_{\rm E}$). Compared to the results for MMSN surface density, more planets have masses comparable to the current Venus and Earth. However, planets that are located near 1.5 au are now a few times more massive than Mars. Increasing the surface density produces more Venus and Earth analogues but less Mars analogues. A larger depletion factor is thus required to produce Mars analogues with the correct mass.

\begin{figure*}[ht]
    \centering
	\includegraphics[width=0.7\textwidth]{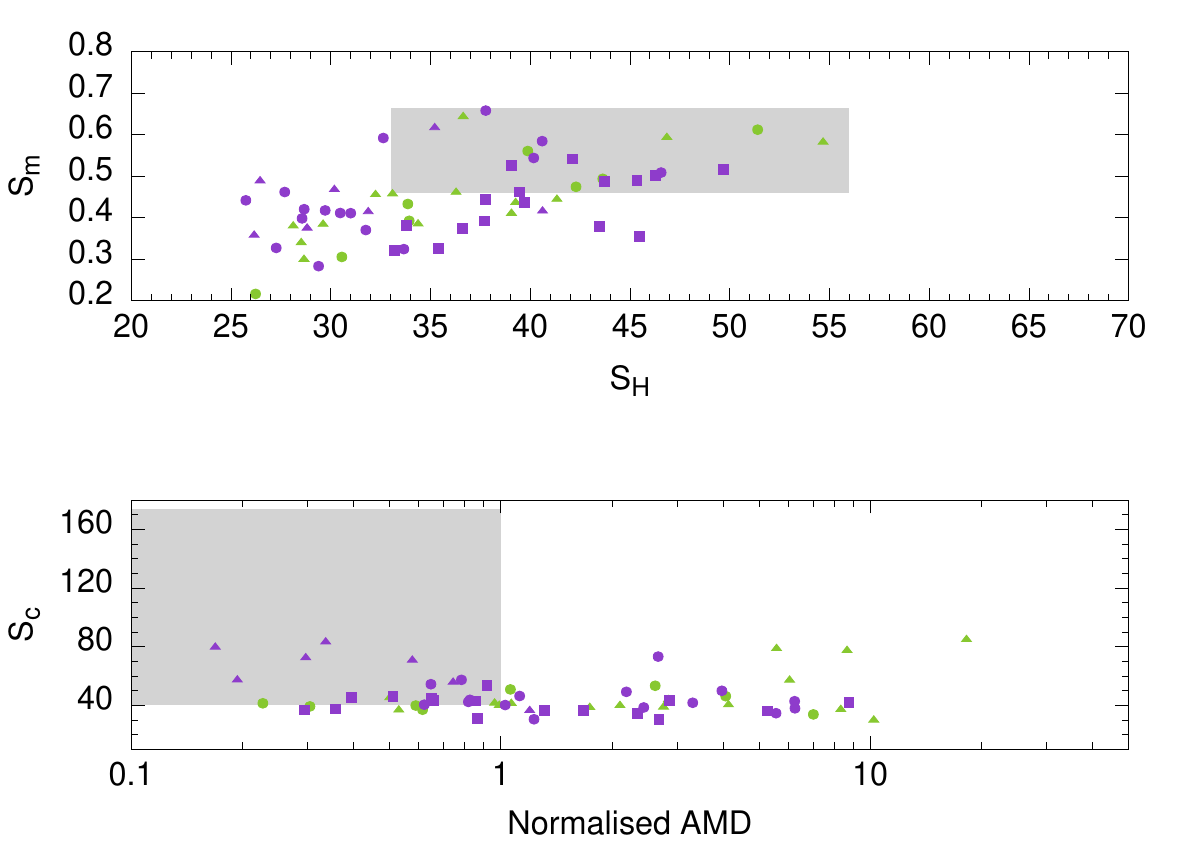}
    \caption{Results of increasing the initial surface density of solids in the protoplanetary disc: Dynamical parameters of planetary systems. Top panel: fraction of total mass in the largest planet $S_m$ versus the mean spacing parameter $S_H$ in units of the mutual Hill sphere. Bottom panel: mass concentration statistic $S_c$ versus the angular momentum deficit (AMD) normalised to the current Solar System value. Green symbols represent simulations with $r_{\rm dep} = 1.5$ au and purple symbols represent $r_{\rm dep} = 1.0$ au. Squares correspond to depletion scale of 50\%, circles to 75\%, and triangles to 95\%. The grey regions represent the current values of the inner Solar System and the 2$\sigma$ range.}
    \label{fig:metrics_extra}
\end{figure*}

As the planets are now more massive on average, the average fraction of the total mass in the largest planet and the average spacing between the planets also increases (top panel of Fig. \ref{fig:metrics_extra}). The results for the concentration parameter $S_c$ and the AMD are similar to the simulations with MMSN surface density (bottom panel of Fig. \ref{fig:metrics_extra}).

In summary, we find with our initial conditions an increase in the initial surface by 1.5 to 2 times the MMSN value is sufficient to reproduce the masses of Venus and Earth but a strong depletion of more than 95\% for $r_{\rm dep} =$ 1.5 au, and more than 75\% up to about 95\% for $r_{\rm dep} =$ 1.0 au is required to reproduce the low mass of Mars.

\section*{Acknowledgements}
Numerical simulations were carried out on the PC cluster at the Center for Computational Astrophysics, National Astronomical Observatory of Japan.


  \bibliographystyle{elsarticle-harv}
  \bibliography{depdisc_doi}

\end{document}